\let\ifarxiv=\iftrue     
\pdfoutput=1

\ifarxiv

\documentclass[12pt,a4paper]{article}

\usepackage{amsmath,amssymb}
\usepackage[nosort]{cite}
\ifnum\pdfoutput=1
\usepackage[bookmarks=true,hyperfigures=true]{hyperref}
\usepackage[final]{graphicx}
\else
\newcommand{\texorpdfstring}[2]{#1}
\usepackage[draft]{graphicx}
\fi
\usepackage[bulletsep]{collect}

\usepackage[a4paper,left=2.5cm,right=2.5cm,top=2.5cm,bottom=2.5cm]{geometry}

\usepackage[font=small,labelfont=bf,width=0.85\textwidth]{caption}

\else
\pdfoutput=1

\documentclass[fdp,a4paper,fleqn]{w-art}
\usepackage{times}
\usepackage[nosort]{cite}
\usepackage{amsmath,amssymb}
\ifnum\pdfoutput=1
\usepackage[bookmarks=true,hyperfigures=true]{hyperref}
\usepackage[final]{graphicx}
\else
\newcommand{\texorpdfstring}[2]{#1}
\usepackage[draft]{graphicx}
\fi

\fi

\ifx\hypersetup\sadfkjashdfkxja\else
\hypersetup{pdftitle={T-Duality, Dual Conformal Symmetry and Integrability for Strings on AdS5xS5}}
\hypersetup{pdfauthor={Niklas Beisert}}
\hypersetup{pdfsubject={}}
\hypersetup{pdfkeywords={T-Duality, Dual Conformal Symmetry, Integrability, Strings on AdS5xS5}}
\hypersetup{plainpages=false}
\hypersetup{bookmarksnumbered=true}
\hypersetup{pdfstartview=FitH}
\hypersetup{pdfpagemode=UseNone}
\hypersetup{colorlinks=false}
\hypersetup{citebordercolor={.5 1 .5}}
\hypersetup{urlbordercolor={.5 1 1}}
\hypersetup{linkbordercolor={1 .7 .7}}
\fi

\makeatletter
\let\old@startsection=\@startsection
\renewcommand{\@startsection}[6]{\old@startsection{#1}{#2}{#3}{#4}{#5}{#6\mathversion{bold}}}
\makeatother

\allowdisplaybreaks[3]

\let\oldPhi=\Phi
\let\oldPsi=\Psi
\let\oldGamma=\Gamma
\let\oldDelta=\Delta
\let\oldSigma=\Sigma
\let\oldLambda=\Lambda
\let\oldTheta=\Theta
\let\oldPi=\Pi
\let\oldOmega=\Omega
\renewcommand{\Phi}{\mathnormal{\oldPhi}}
\renewcommand{\Psi}{\mathnormal{\oldPsi}}
\renewcommand{\Gamma}{\mathnormal{\oldGamma}}
\renewcommand{\Sigma}{\mathnormal{\oldSigma}}
\renewcommand{\Delta}{\mathnormal{\oldDelta}}
\renewcommand{\Theta}{\mathnormal{\oldTheta}}
\renewcommand{\Lambda}{\mathnormal{\oldLambda}}
\renewcommand{\Omega}{\mathnormal{\oldOmega}}
\renewcommand{\Pi}{\mathnormal{\oldPi}}

\newcommand{\gen}[1]{\mathfrak{#1}}

\newcommand{\Integers}{\mathbb{Z}}

\newcommand{\dual}{\mathord{\ast}}
\newcommand{\autom}{\oldOmega}
\newcommand{\superN}{\mathcal{N}}

\ifx\genfrac\sdflkaj

\else

\fi
\newcommand{\sfrac}[2]{{\textstyle\frac{#1}{#2}}}
\newcommand{\half}{\sfrac{1}{2}}

\newcommand{\indup}[1]{_{\mathrm{#1}}}

\newcommand{\lrbrk}[1]{\left(#1\right)}
\newcommand{\bigbrk}[1]{\bigl(#1\bigr)}

\newcommand{\bigcomm}[2]{\big[#1,#2\big]}
\newcommand{\comm}[2]{[#1,#2]}

\newcommand{\alg}[1]{\mathfrak{#1}}
\newcommand{\grp}[1]{\mathrm{#1}}

\newcommand{\nln}{\nonumber\\}
\newcommand{\nl}[1][0pt]{\nonumber\\[#1]&\hspace{-4\arraycolsep}&\mathord{}}

\newcommand{\earel}[1]{\mathrel{}&\hspace{-2\arraycolsep}#1\hspace{-2\arraycolsep}&\mathrel{}}
\newcommand{\eq}{\earel{=}}
\newcommand{\beq}{\begin{equation}}
\newcommand{\eeq}{\end{equation}}

\def\[{\begin{equation}}
\def\]{\end{equation}}
\def\<{\begingroup\ifarxiv\else\arraycolsep1pt\fi\begin{eqnarray}}
\def\>{\end{eqnarray}\endgroup\ignorespaces}

\makeatletter
\def\mr@ignsp#1 {\ifx\:#1\@empty\else #1\expandafter\mr@ignsp\fi}%
\newcommand{\multiref}[1]{\begingroup
\xdef\mr@no@sparg{\expandafter\mr@ignsp#1 \: }%
\def\mr@comma{}%
\@for\mr@refs:=\mr@no@sparg\do{\mr@comma\def\mr@comma{,}\ref{\mr@refs}}%
\endgroup}
\makeatother

\newcommand{\hypref}[2]{\ifx\href\asklfhas #2\else\href{#1}{#2}\fi}

\newcommand{\Figref}[1]{Figure~\multiref{#1}}
\newcommand{\figref}[1]{Fig.~\multiref{#1}}
\renewcommand{\eqref}[1]{(\multiref{#1})}

\ifx\href\asklfhas\newcommand{\href}[2]{#2}\fi
\newcommand{\arxivlink}[1]{\href{http://arxiv.org/abs/#1}{arxiv:#1}}

\begin{document}

\ifarxiv

\begin{flushright}\footnotesize
\texttt{\arxivlink{0903.0609}}\\
\texttt{AEI-2009-027}
\end{flushright}
\vspace{1cm}

\begin{center}%
{\Large\textbf{\mathversion{bold}%
T-Duality, Dual Conformal\\Symmetry and Integrability\\for Strings on $AdS_5\times S^5$}\par}
\vspace{1cm}%

\textsc{Niklas Beisert}\vspace{5mm}%

\textit{Max-Planck-Institut f\"ur Gravitationsphysik\\%
Albert-Einstein-Institut\\%
Am M\"uhlenberg 1, 14476 Potsdam, Germany}\vspace{3mm}%

\texttt{nbeisert@aei.mpg.de}
\par\vspace{1cm}

\textbf{Abstract}\vspace{7mm}

\begin{minipage}{12.7cm}
In recent years two intriguing observations have been made
for $\mathcal{N}=4$ super Yang--Mills theory and 
for superstrings on $AdS_5\times S^5$:
In the planar limit the computation of the spectrum 
is vastly simplified by the apparent integrability of the models.
Furthermore, planar scattering amplitudes of the gauge theory 
display remarkable features which have been attributed
to the appearance of a dual superconformal symmetry. 
Here we review the connection of these two developments 
from the point of view of the classical symmetry
by means of a super-T-self-duality.
In particular, we show explicitly how the charges 
of conformal symmetry and of the integrable structure
are related to the dual ones.
\end{minipage}

\end{center}

\vspace{1cm}
\hrule height 0.75pt
\vspace{1cm}

\else

\DOIsuffix{theDOIsuffix}
\Volume{vv}
\Month{mm}
\Year{yyyy}
\pagespan{1}{}
\Receiveddate{XXXX}
\Reviseddate{XXXX}
\Accepteddate{XXXX}
\Dateposted{XXXX}


\title[T-Duality, Dual Conformal Symmetry and Integrability for Strings on $AdS_5\times S^5$]
{T-Duality, Dual Conformal Symmetry\\and Integrability for Strings on \mathversion{bold}$AdS_5\times S^5$}


\author[N. Beisert]{Niklas Beisert%
  \footnote{E-mail:~\textsf{nbeisert@aei.mpg.de}}}
\address{
\textit{Max-Planck-Institut f\"ur Gravitationsphysik\\%
Albert-Einstein-Institut\\%
Am M\"uhlenberg 1, 14476 Potsdam, Germany}}

\begin{abstract}
In recent years two intriguing observations have been made
for $\mathcal{N}=4$ super Yang--Mills theory and 
for superstrings on $AdS_5\times S^5$:
In the planar limit the computation of the spectrum 
is vastly simplified by the apparent integrability of the models.
Furthermore, planar scattering amplitudes of the gauge theory 
display remarkable features which have been attributed
to the appearance of a dual superconformal symmetry. 
Here we review the connection of these two developments 
from the point of view of the classical symmetry
by means of a super-T-self-duality.
In particular, we show explicitly how the charges 
of conformal symmetry and of the integrable structure
are related to the dual ones.
\end{abstract}

\maketitle

\fi

\section{Introduction}

Two observations of recent years have led to remarkable progress 
in $\superN=4$ supersymmetric gauge theory and in 
IIB string theory on $AdS_5\times S^5$ as well as in their
conjectured duality, the AdS/CFT string/gauge correspondence.
One development is the appearance of integrable structures 
helping dramatically in determining the AdS/CFT spectrum
(we refer the reader to reviews on this subject \cite{Beisert:2004ry,Plefka:2005bk,Arutyunov:2009ga}).
The other observation is that scattering amplitudes in the gauge theory
have a much simpler structure than expected
(see the reviews \cite{Dixon:2008tu,Alday:2008yw}).
Both phenomena have in common that they hold only in the
strict large-$N\indup{c}$ alias the planar limit. 
For a long time the coinciding requirements
have led to speculations that both features may be related. 
Recent works are starting to confirm this idea
and to make the connection more rigorous and concrete. 

Perhaps the first indication of extended symmetries 
for scattering amplitudes was found in \cite{Drummond:2006rz}
where it was argued for the existence of a \emph{dual} conformal symmetry
in addition to the original conformal symmetry. 
This observation helped the four-loop unitarity construction of 
four-gluon scattering \cite{Bern:2006ew}
producing a result for the cusp anomalous dimension 
which is in perfect agreement 
with the prediction based on integrability \cite{Beisert:2006ez}.
Later in \cite{Alday:2007hr} scattering amplitudes 
of the gauge theory were related to certain Wilson loops in the string theory. 
The key step was the proposal of a T-duality transformation on the worldsheet 
coordinates of the string which leaves the (bosonic) action invariant.
The relationship between scattering amplitudes and Wilson loops 
was shown to hold also purely within the perturbative gauge theory setup
\cite{Drummond:2007aua,Brandhuber:2007yx,Drummond:2007cf}
even at higher loops \cite{Drummond:2008aq,Bern:2008ap}.
The conformal symmetry of Wilson loops turns into the
dual conformal symmetry of scattering amplitudes 
\cite{Drummond:2007aua,Drummond:2007au}.
Moreover, the dual symmetry also combines with supersymmetry
into \emph{dual superconformal} symmetry \cite{Drummond:2008vq,Brandhuber:2008pf}.
On the string theory side the above mentioned T-duality transformation 
maps between the original and the dual symmetries.
For the full supersymmetric string action
the standard bosonic T-duality has to be supplemented
by a fermionic T-duality in order to map the
worldsheet action back to itself \cite{Berkovits:2008ic} 
(see also some more recent work \cite{Chandia:2009yv,Adam:2009kt}).

\begin{figure}
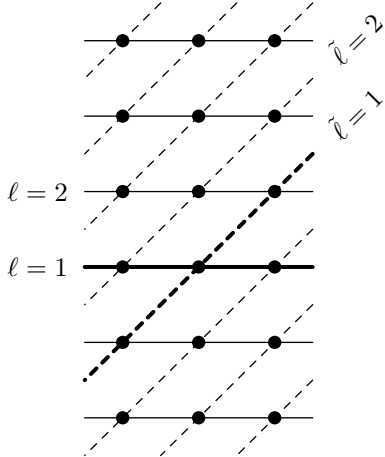
\centering
\ifarxiv\includegraphics{FigTower.mps}\else\includegraphics{FigTowerFDP.mps}\fi
\caption{Embedding of the original and dual superconformal symmetries 
$Q=Y^{(1)}$, $\tilde Q=\tilde Y^{(1)}$ into the integrable structure,
$Y^{(\ell)}$ or $\tilde Y^{(\tilde\ell)}$.}
\label{fig:tower}
\end{figure}

One may wonder what is the closure of the algebra generated 
by the original and the dual symmetries. 
Furthermore, what is the image of the integrable structure
under T-duality? It turns out that both questions have
a common answer \cite{Beisert:2008iq,Berkovits:2008ic,Ricci:2007eq}
(see also \cite{Kluson:2008nn}).
The closure of the algebra is the integrable structure,
and the latter is mapped to itself but in a non-trivial manner.
This is conveniently illustrated in \figref{fig:tower}:
The integrable structure consists of the loop algebra of
superconformal symmetry, i.e.\ infinitely many copies
of the superconformal generators indexed by an integer label.
There are many ways of choosing a closed Lie algebra 
within its loop algebra, and the original and dual symmetries
are two such instances.
Very recently, the integrable structure was shown to apply 
directly to tree-level scattering amplitudes in 
gauge theory \cite{Drummond:2009fd}. 
The loop algebra is quantised to Yangian symmetry.
This symmetry is almost identical to the Yangian symmetry
for one-loop anomalous dimensions \cite{Dolan:2003uh}.
Thus the simplicity of planar scattering amplitudes 
is indeed closely related to the integrability 
of planar AdS/CFT.

This note is a review of the works 
\cite{Alday:2007hr,Berkovits:2008ic,Beisert:2008iq}
outlining T-duality for superstrings on $AdS_5\times S^5$
and how the integrable structure transforms under it. 
We extend the previous works slightly by making the mapping
of the integrable charges more explicit.
We start by reviewing the $AdS_{n}$ coset space sigma model, its
integrable structure as well as its T-duality transformations.
Then we show how T-duality maps the symmetries and the integrable structure
and finally we sketch the extension of the above to the supersymmetric model on $AdS_5\times S^5$.

\section{The \texorpdfstring{$AdS_{n+1}$}{AdS(n+1)} Sigma Model and Integrability}

The anti-de Sitter spacetime $AdS_{n+1}$ is most conveniently viewed
as the symmetric coset space $\grp{SO}(n,2)/\grp{SO}(n,1)$.
Let the algebra $\alg{so}(n,2)$ be spanned by 
the standard conformal generators: 
$\gen{L}_{\mu\nu}$ (Lorentz),
$\gen{D}$ (dilatation),
$\gen{P}_\mu$ (momentum)
and $\gen{K}_\mu$ (special conformal).
These obey the algebra relations%
\footnote{We disregard reality conditions of the algebra
and use this freedom to remove factors of $i$ into the definition
of the generators.} 
\[
\begin{array}[b]{rcl}
\comm{\gen{L}_{\mu\nu}}{\gen{L}_{\rho\sigma}}\eq\eta_{\nu\rho}\gen{L}_{\mu\sigma}\mp \mbox{3 terms},
\\[1ex]
\comm{\gen{P}_\mu}{\gen{K}_\nu}\eq 2\gen{L}_{\mu\nu}+2\eta_{\mu\nu}\gen{D},
\end{array}
\quad
\begin{array}[b]{rcl}
\comm{\gen{L}_{\mu\nu}}{\gen{P}_\rho}\eq\eta_{\nu\rho}\gen{P}_{\mu}-\eta_{\mu\rho}\gen{P}_{\nu},
\\[1ex]
\comm{\gen{L}_{\mu\nu}}{\gen{K}_\rho}\eq\eta_{\nu\rho}\gen{K}_{\mu}-\eta_{\mu\rho}\gen{K}_{\nu},
\end{array}
\quad
\begin{array}[b]{rcl}
\comm{\gen{D}}{\gen{P}_\mu}\eq +\gen{P}_\mu,
\\[1ex]
\comm{\gen{D}}{\gen{K}_\mu}\eq -\gen{K}_\mu.
\end{array}
\]
We embed the denominator algebra $\alg{so}(n,1)$ 
as the invariant space of the $\Integers_2$ automorphism $\autom$ 
of $\alg{so}(n,2)$ defined by
\[
\autom(\gen{L}_{\mu\nu})=\gen{L}_{\mu\nu},\qquad
\autom(\gen{D})=-\gen{D},\qquad
\autom(\gen{P}_\mu)=\gen{K}_\mu,\qquad
\autom(\gen{K}_\mu)=\gen{P}_\mu.
\]

We can formulate the non-linear sigma model on $AdS_{n+1}$ using
the $\grp{SO}(n,2)$-valued field $g$ and its associated Maurer--Cartan form $J=g^{-1}dg$. 
The coset space is implemented through 
the gauge symmetry $g\mapsto gh$ with a $\grp{SO}(n,1)$-valued field $h$
for which $J$ acts as a gauge connection.
Dynamics of the model is governed by a set of invariant equations 
for $J$ and $K=J-\autom(J)$
\[\label{eq:MCEOM}
dJ+J\wedge J=0,\qquad
d\dual K+J\wedge \dual K+\dual K\wedge J=0.
\]
The first expression is the Maurer--Cartan equation following from the
definition of $J$ and the second is the equation of motion which follows 
from the standard non-linear sigma model Lagrangian.

This two-dimensional field theory model turns out to be integrable:
The gauge connection $J$ can be deformed into the Lax connection
\[
A(x)=J+\frac{1}{x^2-1}\,K+\frac{x}{x^2-1}\, \dual K.
\]
Provided that the above equations of motion hold, the Lax
connection is flat, $dA(x)+A(x)\wedge A(x)=0$, 
for all values of the spectral parameter $x$.
Note that the Maurer--Cartan equation for 
the combination $K$ reads $dK+J\wedge K+K\wedge J-K\wedge K=0$.

Now the forms $J$ and $K$ are covariant under gauge transformations
which makes the identification of conserved quantities cumbersome. 
We therefore go to a gauge-invariant frame by conjugating with the group
element $g$.
The field $k=gKg^{-1}$ obeys the equation $dk-k\wedge k=0$ as
well as $d\dual k=0$. In other words, $k$ is the Noether current
leading to the conserved Noether charges 
\[\label{eq:Noether}
Q=\int \dual k.
\]
The Lax connection in the invariant frame 
is obtained by conjugation of the associated
covariant derivative $d+a(x)=g(d+A(x))g^{-1}$,
it reads
\[
a(x)
=\frac{1}{x^2-1}\,k+\frac{x}{x^2-1}\,\dual k
=\frac{x^{-2}}{1-x^{-2}}\,k+\frac{x^{-1}}{1-x^{-2}}\,\dual k.
\]
It is used to construct the higher charges of the integrable model
through its parallel transport,
the so-called monodromy $M(x)$, 
along a curve $\gamma$ on the worldsheet. 
Due to the vanishing of $a(x)$ near $x=\infty$, 
the expansion of $M(x)$ around this point  
\[
M(x)
=\overrightarrow{\mathrm{P}\exp} \int_\gamma a(x)
=\exp\lrbrk{\sum_{n=1}^\infty x^{-n}Y^{(n)}} 
\]
leads to a tower of $n$-local charges $Y^{(n)}$
\<\label{eq:Yn}
Q=Y^{(1)}\eq\int \dual k,
\nln
Y^{(2)}\eq
\half\iint\limits_{\sigma_1<\sigma_2} \comm{\dual k_1}{\dual k_2}
+\int k,
\nln
Y^{(3)}\eq
-\sfrac{1}{6}\iiint\limits_{\sigma_1<\sigma_2<\sigma_3} 
\bigbrk{\bigcomm{\dual k_1}{\comm{\dual k_3}{\dual k_2}}+ \bigcomm{\dual k_3}{\comm{\dual k_1}{\dual k_2}}}
\nl
+\half \iint\limits_{\sigma_1<\sigma_2} \bigbrk{\comm{k_1}{\dual k_2}+\comm{\dual k_1}{k_2}}
+\int \dual k,\quad \ldots
\>
The first of these multi-local charges $Y^{(1)}$ is precisely the Noether charge
$Q$ and thus the integrable structure enhances the Lie algebra $\alg{so}(n,2)$
to an infinite-dimensional algebra.
The higher charges $Y^{(n)}$, $n>1$, are not strictly conserved: 
Shifting the end-points of the curve $\gamma$ in $Y^{(n)}$ 
leads to commutators involving the lower $Y^{(k)}$.

\section{Poincar\'e Coordinates and T-Self-Duality}

The above formulation in terms of a coset still leaves
a large amount of gauge freedom. For $AdS_{n+1}$ there is a
convenient (local) chart of coordinates which fixes the gauge:
It is sufficient to specify the coordinates along the $\gen{P}$ and $\gen{D}$ 
directions of $\grp{SO}(n,2)$ because the $\grp{SO}(n,1)$-directions
$\gen{L}$ and $\gen{P}+\gen{K}$ are unphysical. We can thus choose the group
element to be
\[\label{eq:PoincareG}
g=\exp(X^\mu\gen{P}_\mu)\exp(\Phi\gen{D}).
\]
Here $\Phi$ measures the distance to the boundary
and a slice of constant $\Phi$ is a Minkowski space. 
Consequently we call $(X^\mu,\Phi)$ Poincar\'e coordinates for $AdS_{n+1}$.
The advantage of this chart is that the algebra generated by $\gen{P},\gen{D}$ 
is triangular and the Maurer--Cartan form takes a simple form
\[\label{eq:PoincareJ}
J=g^{-1}dg=J_{\gen{P}}+J_{\gen{D}},\qquad
J_{\gen{P}}=e^{-\Phi}dX^\mu\,\gen{P}_\mu,\quad
J_{\gen{D}}=d\Phi\,\gen{D}.
\]
The Maurer--Cartan equations and the equations of motion read
\[\label{eq:EOMJPD}
\begin{array}[b]{rcl}
0\eq dJ_{\gen{D}}
,\\[1ex]
0\eq
d\dual J_{\gen{D}}
-\half J_{\gen{P}}\wedge \dual\autom(J_{\gen{P}})
-\half \dual\autom(J_{\gen{P}})\wedge J_{\gen{P}}
,
\end{array}
\begin{array}[b]{rcl}
0\eq 
dJ_{\gen{P}}
+J_{\gen{D}}\wedge J_{\gen{P}}
+J_{\gen{P}}\wedge J_{\gen{D}}
,\\[1ex]
0\eq
d\dual J_{\gen{P}}
-J_{\gen{D}}\wedge \dual J_{\gen{P}}
-\dual J_{\gen{P}}\wedge J_{\gen{D}}
.
\end{array}
\]
In fact, the equations of motion for the coordinates $X^\mu,\Phi$ are 
even simpler
\[\label{eq:EOMXPhi}
d(e^{-2\Phi}\dual dX^\mu)=0,
\qquad
d\dual d\Phi-e^{-2\Phi}dX^\mu\wedge \dual dX_\mu=0.
\]

Interestingly, the field $X$ appears in all places only through its derivative $dX$. 
This allows to introduce a set of dual fields $(\tilde X^\mu,\tilde\Phi)$
through the relation 
\[\label{eq:DualXPhi}
d\tilde X^\mu=e^{-2\Phi}\dual dX^\mu,
\qquad
\tilde \Phi=-\Phi.
\]
In fact, the transformation is a combination of a formal T-duality 
on the coordinates $X^\mu$ and a flip of the sign of $\Phi$ 
\cite{Alday:2007hr}.
Note that the transformation between $X$ and $\tilde X$ is non-local
because their relation is defined only via their derivatives. 
It was chosen such that the above equation of motion 
\eqref{eq:EOMXPhi} for $X$ is automatically satisfied, $dd\tilde X=0$.
Conversely, closedness of $dX$ leads to an equation of motion for $\tilde X$. 
Incidentally, it takes precisely the same form as \eqref{eq:EOMXPhi}. 
Likewise the equation of motion for $\tilde\Phi$ matches 
the one of $\Phi$ with all fields replaced by their duals
\[
d(e^{-2\tilde\Phi}\dual d\tilde X^\mu)=0,
\qquad
d\dual d\tilde\Phi-e^{-2\tilde\Phi}d\tilde X^\mu\wedge \dual d\tilde X_\mu=0.
\]
In the first-order formalism involving the Maurer--Cartan form,
the transformation is given through the map
\[\label{eq:Jmap}
\tilde J_{\gen{P}}=\dual J_{\gen{P}},\qquad
\tilde J_{\gen{D}}=-J_{\gen{D}}.
\]
The set of first-order equations \eqref{eq:EOMJPD} is again mapped to itself,
however, the role of Maurer--Cartan equation and equation of motion
for $J_{\gen{P}}$ are interchanged.

\section{T-Self-Duality and Symmetries}

We are thus in the curious situation that the T-duality 
transforms the model to itself, namely it is a T-self-duality 
\cite{Alday:2007hr}.
The model can be expressed through two sets of variables
which incidentally obey the same set of equations.
For all quantities expressed through the original variables
there must therefore exist a quantity expressed through the 
dual variables enjoying the same properties. 
For example, in addition to the Noether charge $Q$ there 
exists a dual Noether charge $\tilde Q$. The associated
symmetry is the so-called dual conformal symmetry.
This symmetry is a $\grp{SO}(n,2)$ group which is
not equivalent to the original conformal $\grp{SO}(n,2)$ symmetry,
although, for example, the Lorentz subgroup of both symmetries coincides.
One also comes to the conclusion 
that the higher integrable charges $Y^{(n)}$ 
lead to dual charges $\tilde Y^{(n)}$. 
An important question is whether these charges are independent 
of the original ones and thus whether the model has two 
coexisting integrable structures. 
Alternatively, there could be a relation between the two 
towers of charges, and if so, what is it precisely?

Let us therefore compare the Lax connection and its dual version
expressed through the original variables via \eqref{eq:Jmap}
\<
A(x)\eq
\frac{1}{x^2-1}\, \bigbrk{
+x^2 J_{\gen{P}}
-\autom(J_{\gen{P}})
+x \dual J_{\gen{P}}
-x \dual \autom(J_{\gen{P}})
+(x^2+1) J_{\gen{D}}
-x\dual J_{\gen{D}}
},
\nln
\tilde A(x)\eq
\frac{1}{x^2-1}\, \bigbrk{
-x \autom(J_{\gen{P}})
+x J_{\gen{P}}
-\dual \autom(J_{\gen{P}})
+x^2 \dual J_{\gen{P}}
-(x^2+1) J_{\gen{D}}
+x\dual J_{\gen{D}}
}.
\>
Looking at the $\gen{D}$-components, it becomes clear
that an algebraic transformation to relate $A(x)$ and $\tilde A(x')$ 
cannot involve the Hodge dual or a change of spectral parameter,
$x\mapsto x'$, but it must act by merely flipping the sign $J_{\gen{D}}\mapsto -J_{\gen{D}}$. 
On the other hand the $\gen{P}$-components imply that the spectral parameter
must be involved in the transformation. 
The first two terms suggest two options for the transformation, 
$J_{\gen{P}}\mapsto x^{-1}J_{\gen{P}}$,
$\autom(J_{\gen{P}})\mapsto x\autom(J_{\gen{P}})$
or 
$J_{\gen{P}}\mapsto -x^{-1}\autom(J_{\gen{P}})$,
$\autom(J_{\gen{P}})\mapsto -xJ_{\gen{P}}$. 
The former does however not lead to the desired result 
for the Hodge dual terms, while the latter one does.
Altogether the transformation can be formulated as an 
$x$-dependent automorphism $\autom_x$
\cite{Beisert:2008iq,Berkovits:2008ic}
\[\label{eq:LaxAuto}
\tilde A(x)=\autom_x(A(x)),\qquad
\autom_x(\gen{X})=(-x)^{\gen{D}}\,\autom(\gen{X})\,(-x)^{-\gen{D}}.
\]
Similarly, the automorphism maps between the parallel transports 
of the Lax connections $A$ and $\tilde A$
\[
\autom_x:
\overrightarrow{\mathrm{P}\exp} \int_\gamma A(x)
\mapsto 
\overrightarrow{\mathrm{P}\exp} \int_\gamma \tilde A(x).
\]
Since the higher integrable charges $Y^{(n)}$ are defined 
in the invariant frame we have to convert this statement 
by conjugation with $g$
\[
\overrightarrow{\mathrm{P}\exp} \int_\gamma A(x)=
g_-^{-1}\lrbrk{\overrightarrow{\mathrm{P}\exp} \int_\gamma a(x)} g_+
=
g_-^{-1} M(x) g_+.
\]
Here $g_\mp$ denote the values of $g$ at the endpoints of the curve $\gamma$. 
The statement is thus
\[
\autom_x\bigbrk{g_-^{-1} M(x) g_+}=\tilde g_-^{-1} \tilde M(x) \tilde g_+.
\]
Using the definition \eqref{eq:PoincareG} of $g$,
the T-duality transformation \eqref{eq:DualXPhi}
and the $\gen{K}$-components of the Noether charges 
(see below)
$\tilde Q_{\gen{K}}=(X_-^\mu-X_+^\mu)\gen{K}_\mu$,
$Q_{\gen{K}}=(\tilde X_-^\mu-\tilde X_+^\mu)\gen{K}_\mu$,
we obtain a useful expression for the relation
between the monodromy matrices
\<
\earel{}
\exp(-x^{-1}\tilde Q_{\gen{K}}) 
\exp(-\tilde X_-^\mu\gen{P}_\mu) \tilde M(x) \exp(+\tilde X_-^\mu\gen{P}_\mu)\nln
\eq
\autom_x\bigbrk{
\exp(-X_+^\mu\gen{P}_\mu) M(x) \exp(+X_+^\mu\gen{P}_\mu)
\exp(-x^{-1}Q_{\gen{K}})
}.
\>

Conveniently the conjugations by $\exp(X_+^\mu\gen{P}_\mu)$
and $\exp(\tilde X_-^\mu\gen{P}_\mu)$
lead to only finitely many terms due to the nilpotency of $\gen{P}$.
As a first step, let us expand all exponents to linear order in the
exponent
\[
\sum_{n=1}^\infty x^{-n}\tilde Y^{(n)}
\simeq
x^{-1}\tilde Q_{\gen{K}}+
\sum_{n=1}^\infty x^{-n} \autom_x(Y^{(n)})
-x^{-1}\autom_x(Q_{\gen{K}}).
\]
These expressions are exact up to commutators involving
the lower charges as well as $X_+^\mu\gen{P}_\mu$ or $\tilde X_-^\mu\gen{P}_\mu$.
When we split this into components we find the relations
($n\geq 1$)
\[
\tilde Y^{(n+1)}_{\gen{K}} \simeq-\autom(Y^{(n)}_{\gen{P}}),
\quad
\tilde Y^{(n)}_{\gen{P}}\simeq-\autom(Y^{(n+1)}_{\gen{K}}),
\quad
\tilde Y^{(n)}_{\gen{D}}\simeq-Y^{(n)}_{\gen{D}},
\quad
\tilde Y^{(n)}_{\gen{L}}\simeq Y^{(n)}_{\gen{L}}.
\]
We can also write down the first few relations with the omitted commutators explicitly: 
\<\label{eq:YnRel}
\tilde Q_{\gen{L}}+\tilde Q_{\gen{D}}
-\comm{\tilde X_-^\mu\gen{P}_\mu}{\tilde Q_{\gen{K}}}
\eq
\autom\bigbrk{
Q_{\gen{L}}+Q_{\gen{D}}
-\comm{X_+^\mu\gen{P}_\mu}{Q_{\gen{K}}}
},
\\\nonumber
\tilde Q_{\gen{P}}
-\bigcomm{\tilde X_-^\mu\gen{P}_\mu}{\tilde Q_{\gen{L}}+\tilde Q_{\gen{D}}-\half \comm{\tilde X_-^\mu\gen{P}_\mu}{\tilde Q_{\gen{K}}}}
\eq
\autom\bigbrk{
-Y^{(2)}_{\gen{K}} 
+\half \bigcomm{Q_{\gen{L}}+Q_{\gen{D}}-\comm{X_+^\mu\gen{P}_\mu}{Q_{\gen{K}}}}{Q_{\gen{K}}}
},
\\\nonumber
\tilde Y^{(2)}_{\gen{K}}
+\half \bigcomm{\tilde Q_{\gen{L}}+\tilde Q_{\gen{D}}-\comm{\tilde X_-^\mu\gen{P}_\mu}{\tilde Q_{\gen{K}}}}{\tilde Q_{\gen{K}}}
\eq
\autom\bigbrk{
-Q_{\gen{P}}
+\bigcomm{X_+^\mu\gen{P}_\mu}{Q_{\gen{L}}+Q_{\gen{D}}-\half\comm{X_+^\mu\gen{P}_\mu}{Q_{\gen{K}}}}
}.\hspace*{-6pt}
\>

Let us see in practice, how the duality of charges works. 
First we work out the components of the Noether current $k=gKg^{-1}$
from \eqref{eq:PoincareG,eq:PoincareJ}
\[\begin{array}{rcl}
k_{\gen{K}}\eq -e^{-2\Phi}dX^\mu\gen{K}_\mu,
\\
k_{\gen{D}}\eq 2(d\Phi-e^{-2\Phi}X_\mu dX^{\mu})\gen{D},
\\
k_{\gen{L}}\eq -2e^{-2\Phi} X^\mu dX^\nu\gen{L}_{\mu\nu},
\\
k_{\gen{P}}\eq (dX^\mu-2X^\mu d\Phi-e^{-2\Phi}X^2 dX^\mu+2e^{-2\Phi}X^\mu X_\nu dX^\nu )\gen{P}_{\mu}.
\end{array}
\]
Their Hodge duals will be written using the dual coordinates as far as possible
\[\begin{array}{rcl}
\dual k_{\gen{K}}\eq -d\tilde X^\mu\gen{K}_\mu,
\\
\dual k_{\gen{D}}\eq 2(\dual d\Phi-X_\mu d\tilde X^{\mu})\gen{D},
\\
\dual k_{\gen{L}}\eq -2 X^\mu d\tilde X^\nu\gen{L}_{\mu\nu},
\\
\dual k_{\gen{P}}\eq (e^{2\Phi}d\tilde X^\mu-2X^\mu \dual d\Phi-X^2d\tilde X^\mu+2X^\mu X_\nu d\tilde X^\nu )\gen{P}_{\mu}.
\end{array}
\]
The $\gen{K}$-components of the Noether charges \eqref{eq:Noether}
read
\[
Q_{\gen{K}}=\int \dual k_{\gen{K}}=-\int d\tilde X^\mu\gen{K}_\mu=
(\tilde X_-^\mu-\tilde X_+^\mu)\gen{K}_\mu,
\qquad
\tilde Q_{\gen{K}}=\ldots=(X_-^\mu-X_+^\mu)\gen{K}_\mu.
\]
These two quantities are independent, and the 
relationship between the monodromies respects this.
Next we consider their $\gen{L}$-components
\[
Q_{\gen{L}}=-2\int X^\mu d\tilde X^\nu\gen{L}_{\mu\nu},
\qquad
\tilde Q_{\gen{L}}
=-2\int \tilde X^\mu dX^\nu\gen{L}_{\mu\nu}.
\]
Upon partial integration we recover the expression for 
$Q_{\gen{L}}$ in $\dual Q_{\gen{L}}$ up to some boundary terms
\[
\tilde Q_{\gen{L}}
=
Q_{\gen{L}}
+2(\tilde X_-^\mu X_-^\nu-\tilde X_+^\mu X_+^\nu )\gen{L}_{\mu\nu}
=
Q_{\gen{L}}
+2\bigbrk{(\tilde X_-^\mu-\tilde X_+^\mu )X_+^\nu+\tilde X_-^\mu (X_-^\nu-X_+^\nu) }\gen{L}_{\mu\nu}.
\]
This is precisely the $\gen{L}$-component of the first equation in \eqref{eq:YnRel},
similarly for $\gen{D}$.
It is also interesting to consider the bi-local charge $Y^{(2)}_{\gen{K}}$
\cite{Berkovits:2008ic}
\[
Y^{(2)}_{\gen{K}}
=
\half\iint\limits_{\sigma_1<\sigma_2} \comm{\dual k_{1,\gen{K}}}{\dual k_{2,\gen{L}+\gen{D}}}
+\half\iint\limits_{\sigma_1<\sigma_2} \comm{\dual k_{1,\gen{L}+\gen{D}}}{\dual k_{2,\gen{K}}}
+\int k_{\gen{K}}.
\]
Now because $\dual k_{\gen{K}}$ is a total derivative 
the double integral collapses to a single one
\[
Y^{(2)}_{\gen{K}}
=
\half\bigcomm{(\tilde X_-^\mu+\tilde X_+^\mu) \gen{K}_\mu}{Q_{\gen{L}}+Q_{\gen{D}}}
-\int\comm{\tilde X^\mu \gen{K}_\mu}{\dual k_{\gen{L}+\gen{D}}}
+\int k_{\gen{K}}.
\]
After substituting the remaining expressions and some partial integrations
one finds the expression for $\autom(\tilde Q_{\gen{P}})$ plus some boundary terms
\<
Y^{(2)}_{\gen{K}}
\eq
-\autom(\tilde Q_{\gen{P}})
+\half\bigcomm{(\tilde X_-^\mu+\tilde X_+^\mu) \gen{K}_\mu}{Q_{\gen{L}}+Q_{\gen{D}}}
\nl
+2(X_+\cdot\tilde X_+) \tilde X_+^\mu  \gen{K}_\mu
+2(X_-\cdot\tilde X_-) \tilde X_-^\mu  \gen{K}_\mu
-\tilde X_+^2 X_+^\mu \gen{K}_\mu
+\tilde X_-^2 X_-^\mu \gen{K}_\mu.
\>
Some elementary operations later
one recovers precisely the second relation in \eqref{eq:YnRel}.

\section{Super-T-Self-Duality and Integrability}

An analog construction exists for the 
superstring on $AdS_5\times S^5$ 
or, equivalently, the sigma model
on the coset space $\grp{PSU}(2,2|4)/\grp{Sp}(1,1)\times\grp{Sp}(2)$
coupled to worldsheet gravity.
The coset construction is based on a $\Integers_4$ grading 
which allows to split the Maurer--Cartan into four components
\[
J=J_0+J_2+J_1+J_{-1},\qquad
\autom(J_n)=i^n J_n,\quad
J_n= \frac{1}{4}\sum_{k=0}^3 i^{-nk}\autom^{\circ k}(J).
\]
By introducing the combination 
\[
\dual K=2\dual J_2-J_1+J_{-1}
\]
we can write the Maurer--Cartan equations and the equations
of motion in precisely the same way as in \eqref{eq:MCEOM}.
Note, however, that splitting these equations into their 
$\Integers_4$ components yields a more complicated set of equations than before.
These equations can again be cast into the form of a flatness condition 
for a Lax connection \cite{Bena:2003wd}
\[
A(z)=J_0+\half (z^2+z^{-2})J_2+\half (-z^2+z^{-2})\dual J_2+z J_1 +z^{-1} J_{-1}.
\]
Note that the bosonic part consisting of the first three terms 
is the same as above if the spectral parameters are identified as 
\cite{Beisert:2005bm}
\[\label{eq:xvsz}
z^2=\frac{x-1}{x+1}\,,\qquad
x=\frac{1+z^2}{1-z^2}\,.
\]

\begin{figure}
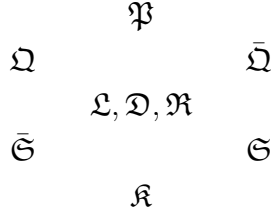
\centering
$\begin{array}{ccccc}
 && \gen{P} && 
\\[0.5ex]
 \gen{Q} && && \gen{\bar Q}
\\[0.5ex]
 && \gen{L,D,R} && 
\\[0.5ex]
 \gen{\bar S}&& && \gen{S}
\\[0.5ex]
 && \gen{K} &&
\end{array}$
\caption{Structure of the superconformal algebra $\alg{psu}(2,2|4)$.
The generators are arranged in the plane according to their charges 
under two Cartan generators $(\gen{B},\gen{D})$
which are conserved in commutators.}
\label{fig:superconf}
\end{figure}

The $\alg{psu}(2,2|4)$ algebra is spanned by the above
$\alg{so}(4,2)$ conformal generators $\gen{L}$, $\gen{D}$, $\gen{P}$, $\gen{K}$,
by the internal $\alg{so}(6)$ generators $\gen{R}$ 
as well as the supercharges $\gen{Q}$, $\gen{\bar Q}$, $\gen{S}$, $\gen{\bar S}$.
The structure of the algebraic relations can conveniently be 
sketched as in \figref{fig:superconf}.

\[
\autom(\gen{Q})\sim i\gen{S},\qquad
\autom(\gen{S})\sim i\gen{Q},\qquad
\autom(\gen{\bar Q})\sim i\gen{\bar S},\qquad
\autom(\gen{\bar S})\sim i\gen{\bar Q}.
\]
In addition to the bosonic gauge group $\grp{Sp}(1,1)\times\grp{Sp}(2)$
with grading $0$,
the model has a local fermionic kappa symmetry
affecting predominantly the components with grading $\pm 1$. 
These local symmetries can be gauge fixed in many different useful ways.
For our purposes it is again advisable to choose them such that 
as many components of the Maurer--Cartan form as possible become trivial.
Again we would like to eliminate the components corresponding 
to conformal boosts, $J_{\gen{K}}=0$. In addition we can 
eliminate half of the fermionic components. 
One option is to eliminate the components corresponding to all superconformal boosts,
$J_{\gen{S}}=J_{\gen{\bar S}}=0$. This is achieved by 
choosing the group element to be generated by 
$\gen{P}$, $\gen{Q}$, $\gen{\bar Q}$, $\gen{D}$, $\gen{R}$
\cite{Kallosh:1998zx,Kallosh:1998ji,Pesando:1998fv}.
\Figref{fig:superconf} shows that commutators of the generators close onto
the subset, and we can understand this choice as upper triangular matrices.
By the same logic we can instead eliminate 
$J_{\gen{S}}=J_{\gen{\bar Q}}=0$ by the alternative
choice of group element generated by 
$\gen{P}$, $\gen{Q}$, $\gen{\bar S}$, $\gen{D}$, $\gen{R}$
\cite{Roiban:2000yy}.%
\footnote{This gauge necessarily requires complexification 
of the coordinates.}

Both gauges have in common that they make $4$ bosonic and $8$ fermionic
coordinate fields appear only through their derivatives. 
Therefore one can apply a T-duality transformation to all of these fields. 
It is a combination of a bosonic T-duality similar to \eqref{eq:DualXPhi}
and a so-called fermionic T-duality acting on the fermionic fields
\cite{Berkovits:2008ic}.%
\footnote{Note that T-duality in $n=4$ bosonic variables
induces a shift of the dilaton which is 
cancelled precisely by T-duality in $2n=8$ fermionic variables
leading to a quantum mechanically exact self-duality.}
In contradistinction to the purely bosonic case, 
the T-duality transformation does not map the equations of motion 
into themselves. It rather maps between the equations in 
the two gauges discussed above.

The formulation of the T-duality transformation for the fields 
is somewhat complicated and it depends on the precise choice of gauge.
In fact, it is much simpler to state the resulting transformation 
in terms of the Maurer--Cartan form analogously to \eqref{eq:Jmap}
\[\label{eq:superJmap}
\tilde J_{\gen{P}}=\dual J_{\gen{P}}
,\quad
\tilde J_{\gen{D}}= -J_{\gen{D}}
,\quad
\tilde J_{\gen{Q}}=iJ_{\gen{Q}}
,\quad
\tilde J_{\gen{\bar S}}=\autom(J_{\gen{\bar Q}})
,\quad
\tilde J_{\gen{R}}= \autom(J_{\gen{R}}).
\]
The statement is that 
when taking the full set of equations for $J$
and restricting them to the above two gauge choices, 
T-duality will map between the two sets. 
Instead of proving the statement, we will show 
that the Lax connections are related by an automorphism as 
in the bosonic case, which also proves the equivalence
of the integrable structure and its dual.
We first split the Lax connection into bosonic and fermionic 
components, $A(z)=A\indup{B}(z)+A\indup{F}(z)$
which must transform separately.
The transformation for the bosonic components works 
as in the purely bosonic case \eqref{eq:LaxAuto} 
after identifying $x$ and $z$ according to \eqref{eq:xvsz}.
We are left with comparing the fermionic components 
of the Lax connection in one gauge and its dual in a different gauge
but expressed through the first set of variables via \eqref{eq:superJmap}
\<
A\indup{F}(z)\eq
\half (z+z^{-1})\bigbrk{J_{\gen{Q}}+J_{\gen{\bar Q}}}
+\half (z-z^{-1})\bigbrk{-i\autom(J_{\gen{Q}})-i\autom(J_{\gen{\bar Q}})}
,
\\
\tilde A\indup{F}(z)\eq
\half (z+z^{-1})\bigbrk{+iJ_{\gen{Q}} +\autom(J_{\gen{\bar Q}})}
+\half (z-z^{-1})\bigbrk{\autom(J_{\gen{Q}})+iJ_{\gen{\bar Q}}}
.
\>
From the bosonic part we know that the automorphism involves the
$\Integers_4$ transformation $\autom$. Consequently, here we
are forced to interchange the $J_{\gen{Q}}$ terms 
which implies a $z$-dependent factor for $J_{\gen{Q}}$.
Conversely the $J_{\gen{\bar Q}}$ terms must stay in place
and the automorphism should be independent of $z$.
Altogether this is achieved by the following transformation
\cite{Beisert:2008iq}
\[
\tilde A(z)=\autom_z(A(z)),\qquad
\autom_z(\gen{X})=
\lrbrk{-\frac{1+z^2}{1-z^2}}^{\gen{D}+\gen{B}}\autom\bigbrk{\gen{X}}\lrbrk{-\frac{1-z^2}{1+z^2}}^{\gen{D}+\gen{B}}.
\]
Here $\gen{B}$ generates the $\alg{u}(1)$ automorphism of $\alg{psu}(2,2|4)$
which acts exclusively on the fermionic generators $\gen{Q}$, $\gen{\bar Q}$,
$\gen{S}$, $\gen{\bar S}$.
More concretely $\gen{B}$ is defined such that, 
\<
\comm{\gen{D}+\gen{B}}{(\gen{P},\gen{Q})}\eq +(\gen{P},\gen{Q}),\nln
\comm{\gen{D}+\gen{B}}{(\gen{D},\gen{L},\gen{R},\gen{\bar Q},\gen{\bar S})}\eq 0,\nln
\comm{\gen{D}+\gen{B}}{(\gen{K},\gen{S})}\eq -(\gen{K},\gen{S}).
\>
Clearly the automorphism $\autom_z$ is compatible with 
$\autom_x$ for bosonic generators.
The resulting mapping of non-local charges is analogous to 
the bosonic case; up to commutator terms, it is depicted in \figref{fig:chargemap}.
\begin{figure}\centering
$\begin{array}{ccccc}
 && Y^{(r)}_{\gen{P}}\simeq-\tilde Y^{(r-1)}_{\gen{K}} && 
\\[0.5ex]
 Y^{(r)}_{\gen{Q}}\simeq- \tilde Y^{(r-1)}_{\gen{S}} && && Y^{(r)}_{\gen{\bar Q}}\simeq \tilde Y^{(r\pm0)}_{\gen{\bar S}}
\\[0.5ex]
 && Y^{(r)}_{\gen{L,D,R}}\simeq \pm \tilde Y^{(r\pm0)}_{\gen{L,D,R}} && 
\\[0.5ex]
 Y^{(r)}_{\gen{\bar S}}\simeq \tilde Y^{(r\pm0)}_{\gen{\bar Q}} && && Y^{(r)}_{\gen{S}}\simeq -\tilde Y^{(r+1)}_{\gen{Q}}
\\[0.5ex]
 && Y^{(r)}_{\gen{K}}\simeq -\tilde Y^{(r+1)}_{\gen{P}} &&
\end{array}$
\caption{Mapping between non-local charges for classical superstrings on $AdS_5\times S^5$ and their duals
(up to commutators).}
\label{fig:chargemap}
\end{figure}

\section*{Acknowledgements}

I thank Ricardo Ricci, Arkady Tseytlin and Martin Wolf
for the fruitful collaboration in the article \cite{Beisert:2008iq}
on which this work is based to a large extent.
I acknowledge hospitality at the DESY Workshop on Applied 2d Sigma Models (Hamburg, 2008),
the KITP Workshop on Fundamental Aspects of Superstring Theory (Santa Barbara, 2009)
and the RTN Forces-Universe Workshop (Varna, 2008) 
where this and related work has been presented. 
This research was supported in part by the US National Science Foundation 
under Grant No.\ PHY05-51164.

\bibliography{DualQ}

\ifarxiv
\bibliographystyle{nb}
\else
\bibliographystyle{fdp}
\fi

\end{document}